\documentclass[twocolumn]{aastex62}
\usepackage{amssymb}
\usepackage{natbib}

\newcommand{\specialcell}[2][c]{%
  \begin{tabular}[#1]{@{}c@{}}#2\end{tabular}}

\graphicspath{{./}{figures/}}

\received{07-02-2022}
\accepted{yy-yy-2022}

\shorttitle{}
\shortauthors{}

\begin{document}

\title{ALMA as a Redshift Machine: \\ Using [CII] to Efficiently Confirm Galaxies in the Epoch of Reionization}

\author[0000-0001-9746-0924]{Sander Schouws}
\affil{Leiden Observatory, Leiden University, NL-2300 RA Leiden, Netherlands}

\author[0000-0002-4989-2471]{Rychard Bouwens}
\affil{Leiden Observatory, Leiden University, NL-2300 RA Leiden, Netherlands}

\author[0000-0001-7768-5309]{Renske Smit}
\affil{Astrophysics Research Institute, Liverpool John Moores University, 146 Brownlow Hill, Liverpool L3 5RF, United Kingdom}

\author[0000-0001-6586-8845]{Jacqueline Hodge}
\affil{Leiden Observatory, Leiden University, NL-2300 RA Leiden, Netherlands}

\author[0000-0001-7768-5309]{Mauro Stefanon}
\affil{Leiden Observatory, Leiden University, NL-2300 RA Leiden, Netherlands}

\author[0000-0002-2057-5376]{Joris Witstok}
\affil{Kavli Institute for Cosmology, University of Cambridge, Madingley Road, Cambridge CB3 0HA, UK}
\affil{Cavendish Laboratory, University of Cambridge, 19 JJ Thomson Avenue, Cambridge CB3 0HE, UK}

\author[0000-0000-0000-0000]{Juli\"ette Hilhorst}
\affil{Leiden Observatory, Leiden University, NL-2300 RA Leiden, Netherlands}

\author[0000-0002-2057-5376]{Ivo Labb\'e}
\affil{Centre for Astrophysics and SuperComputing, Swinburne, University of Technology, Hawthorn, Victoria, 3122, Australia}


\author[0000-0002-4205-9567]{Hiddo Algera}
\affil{Hiroshima Astrophysical Science
Center, Hiroshima University, 1-3-1 Kagamiyama, Higashi-Hiroshima,
Hiroshima 739-8526, Japan}

\author[0000-0002-3952-8588]{Leindert Boogaard}
\affil{Max Planck Institute for Astronomy, K\"{o}nigstuhl 17, 69117 Heidelberg, Germany}

\author[0000-0003-0695-4414]{Michael Maseda}
\affil{Department of Astronomy, University of Wisconsin-Madison, 475 N. Charter St., Madison, WI 53706 USA}

\author[0000-0001-5851-6649]{Pascal Oesch}
\affil{Departement d’Astronomie, Universit\'e de Gen\'eeve, 51 Ch. des Maillettes, CH-1290 Versoix, Switzerland}
\affil{International Associate, Cosmic Dawn Center (DAWN), Niels Bohr Institute, University of Copenhagen and DTU-Space, Technical University of Denmark}

\author[0000-0000-0000-0000]{Huub R\"ottgering}
\affil{Leiden Observatory, Leiden University, NL-2300 RA Leiden, Netherlands}

\author[0000-0002-4389-832X]{Paul van der Werf}
\affil{Leiden Observatory, Leiden University, NL-2300 RA Leiden, Netherlands}

\begin{abstract}

The [CII]$_{158\mu m}$ line has long been proposed as a promising line to spectroscopically confirm galaxies in the epoch of reionization.  In this paper we present the results of new ALMA observations spectral scanning for [CII] in six particularly luminous Lyman Break Galaxies at $z\sim7$.  The six sources were drawn from a sample of bright $z\sim7$ galaxies identified using the wide-area optical, near-IR, and \textit{Spitzer}/IRAC data over the COSMOS/UltraVISTA field and were targeted on the basis of tight constraints on their redshifts from their IRAC [3.6]-[4.5] colors.  We detect significant ($>9\sigma$) [CII] lines in three  of our six targets ($50\%$) co-spatial with the rest-$UV$ emission from the ground/space-based near-IR imaging.  The luminosities of the [CII] lines lie in the range $5.6$ to $8.8\times10^{8}L_{\odot}$, consistent with the local [CII]-SFR relation.  Meanwhile, their [CII]/$L_{IR}\sim1-3\times10^{-3}$ ratios are slightly elevated compared to local (U)LIRGS. This could be due to lower dust-to-gas or dust-to-metal ratios. We also find that our sources display a large kinematic diversity, with one source showing signs of rotation, one source a likely major merger and one dispersion dominated source that might contain a bright star-forming clump.  Our results highlight the effectiveness of spectral scans with ALMA in spectroscopically confirming luminous galaxies in the epoch of reionization, something that is being be applied on a significantly larger sample in the on-going REBELS large program.

\end{abstract}


\section{Introduction} \label{sec:intro}

Exploring the build-up of galaxies in the $z>6.5$ universe just a few hundred million years after the Big Bang is a key frontier in extragalactic astronomy. Though 100s of UV-bright galaxy candidates at redshifts $z>6.5$ \citep[e.g.,][]{mclure2013,Bouwens_2015,Finkelstein_2015,Ishigaki_2018,Bouwens_2021,Harikane2021} are known, characterization of their physical properties has been difficult. Deriving these properties from optical and near-IR photometry is complicated by uncertainties in the redshifts \citep[e.g.,][]{Bouwens_LP,Robertson_2021}, dust extinction \citep[e.g.,][]{Fudamoto2020arXiv200410760F,Schouws_2021,Bowler_2021}, and rest-frame optical nebular emission line properties  \citep[e.g.,][]{Smit_2015,Endsley_2021,stefanon_2021a} of $z>6$ galaxies.

Fortunately, with ALMA, it is possible to make great progress on the characterization of galaxies at $z>6$ \citep[e.g.,][]{Hodge2020arXiv200400934H,Robertson_2021}. The [CII]158$\mu$m line is especially interesting as it is the strongest cooling line of warm gas ($T < 10^4$ K) in galaxies. This line has already been detected in a significant number of galaxies out to $z\sim 7$-8 \citep[e.g., ][]{Walter_2009,Wagg_2010,Riechers_2014,Willott_2015,Capak2015_Natur.522..455C,Maiolino_2015_10.1093/mnras/stv1194,Pentericci_2016,Knudsen_2016,Bradac_2017,Smit_2018Natur.553..178S,Matthee_2017,Matthee_2019,harikane_2020,carniani_2020,lefevre_2020,bethermin2020A&A...643A...2B,venemans_2020,Fudamoto_2021}.  In addition to the immediate utility of [CII] to spectroscopically confirm galaxies in the reionization era and obtain a precise measurement of their redshift, the strength of this line makes it one of the prime features for studying the kinematics of high-z galaxies \citep[e.g.,][]{Smit_2018Natur.553..178S,Hashimoto_2019_10.1093/pasj/psz049,jones_2021}. Dynamical masses are particularly interesting in that they give some handle on the masses of galaxies, which can be challenging to do from the photometry alone \citep[e.g.,][]{schaerer_2009}.

Despite the great utility of [CII] and the huge interest from the community, only a modest number of $z>6.2$ galaxies had been found to show [CII] emission in the first six years of ALMA operations \citep[e.g.,][]{Maiolino_2015_10.1093/mnras/stv1194,Pentericci_2016,Bradac_2017,Matthee_2017,Smit_2018Natur.553..178S,Carniani_2018,Hashimoto_2019_10.1093/pasj/psz049}.  Additionally, even in cases where the [CII] line was detected, only modest luminosities were found, i.e., $L_{[CII]}$$\lesssim\,$2$\times$10$^{8}$ $L_{\odot}$.  One potentially important factor in the limited success of earlier probes for [CII] at $z>6$ may have been the almost exclusive targeting of sources with redshifts from the Ly$\alpha$ emission line.  At lower redshifts at least, Ly$\alpha$ emission seems to be much more prevalent in lower mass galaxies than it is in higher mass galaxies \citep[e.g.,][]{Stark_2010}.  Additionally, Ly$\alpha$ emitters have been found to be systematically low in their [CII] luminosity-to-SFR ratios \citep{Harikane_2018,Matthee_2019}.  Both factors would have caused early ALMA observations to miss those galaxies with the brightest [CII] lines.

In a cycle 4 pilot, \citet{Smit_2018Natur.553..178S} demonstrated the effectiveness of spectral scans for the [CII] line in $z>6$ galaxies which are particularly luminous and which also have well constrained photometric redshifts.  One aspect of the $z\sim7$ galaxies from \citet{Smit_2018Natur.553..178S} that allowed for particularly tight constraints on their redshifts were the high EWs of their strong [OIII]+H$\beta$ emission lines in the {\it Spitzer}/IRAC bands.  This is due to the particular sensitivity of the {\it Spitzer}/IRAC [3.6]$-$[4.5] colors to redshift of galaxies for high-EW [OIII]+H$\beta$ emitters \citep{Smit_2015}.  Remarkably, despite just $\sim$1 hour of observations becoming available on these targets, the results were nevertheless striking, with 6-8$\sigma$ [CII] lines found in two targeted sources, with redshifts $z=6.8075$ and $z=6.8540$.  Additionally, the [CII] luminosities of the two sources were relatively high, being brighter in [CII] than all ALMA non-QSO targets by factors of $\sim$2-20.  

Encouraged by the high efficiency of the \citet{Smit_2018Natur.553..178S} spectral scan results, we successfully proposed for similar observations for 6 other luminous $z\sim7$ galaxies (2018.1.00085.S: PI Schouws) to add to the results available from the \citet{Smit_2018Natur.553..178S} program and further refine the spectral scan strategy.  The purpose of this paper is to present results from this second pilot.  Results from this pilot program and an earlier program from \citet{Smit_2018Natur.553..178S} served as the  motivation for the Reionization Era Bright Emission Line Survey (REBELS) ALMA large program \citep{Bouwens_LP} in cycle 7.  The REBELS program significantly expanded the strategy employed in these pilot programs to a much larger set of targets, while extending the scan strategy to even higher redshift.

The paper is structured as follows.  In  \textsection 2, we detail the selection of targets for this second pilot program, while also describing the set up and reduction of our spectral scan observations.  In \textsection 3, we describe the [CII] line search and present our new [CII] detections. We place our detections on the [CII]-SFR relation and examine the [CII]/$L_{IR}$ of our sources. We conclude the Section by examining their kinematics.  In \textsection 4, we discuss the prospect of deploying the described spectral scan observations to a much larger set of $z>6$ galaxies.  Finally in \textsection 5, we summarize our conclusions. 

Throughout this paper we assume a standard cosmology with $H_0=70$ km s$^{-1}$ Mpc$^{-1}$, $\Omega_m=0.3$ and $\Omega_{\Lambda}=0.7$. Magnitudes are presented in the AB system \citep{oke_gunn_1983ApJ...266..713O}. For star formation rates and stellar masses we adopt a Chabrier IMF \citep{Chabrier_2003}. Error-bars indicate the $68\%$ confidence interval unless specified otherwise.

\section{High-Redshift Targets and ALMA Observations}

\subsection{UltraVISTA Search Field and Photometry}

Our source selection is based on ultradeep near-infrared imaging over the
COSMOS field \citep{Scoville_2007} from the third data release (DR3)
of UltraVISTA \citep{McCracken_2012A&A...544A.156M}. UltraVISTA provides imaging
covering 1.4 square degrees \citep{McCracken_2012A&A...544A.156M} in the Y, J, H and
Ks filters to $\sim$24-25 mag (5$\sigma$), with DR3 achieving
fainter limits over 0.7 square degrees in 4 ultradeep strips.  The DR3
contains all data taken between December 2009 and July 2014 and
reaches $Y = 26.2$, $J = 26.0$, $H = 25.8$, $K = 25.5$ mag (5$\sigma$ in 1.2$''$-diameter apertures). The nominal depth we measure
in the Y, J, H, and Ks bands for the UltraVISTA DR3 release is
$\sim$0.1-0.2 mag, $\sim$0.6 mag, $\sim$0.8 mag, and $\sim$0.2 mag,
respectively, deeper than in UltraVISTA DR2 release.

The optical data we use consists of CFHT/Omegacam in g, r, i, z \citep{Erben_2009A&A...493.1197E} from the Canada-France-Hawaii Legacy Survey (CFHTLS) and
Subaru/SuprimeCam in B$_J$+V$_J$+g+r+i+z imaging \citep{Taniguchi_2007}.  This analysis uses Spitzer/IRAC 3.6$\mu$m and 4.5$\mu$m
observations from S-COSMOS \citep{Sanders_2007}, the Spitzer Extended
Deep Survey \cite[SEDS:][]{Ashby_2013}, the Spitzer-Cosmic Assembly
Near-Infrared Deep Extragalactic Survey \citep[S-CANDELS:][]{Ashby_2015}, Spitzer Large Area Survey with Hyper-Suprime-Cam (SPLASH,
PI: Peter Capak), and the Spitzer Matching survey of the UltraVISTA
ultra-deep Stripes \citep[SMUVS, PI: K. Caputi:][]{Ashby_2018}. Compared to the original S-COSMOS IRAC data,
SPLASH provides a large improvement in depth over nearly the whole
UltraVISTA area, covering the central 1.2 square degree COSMOS field
to 25.5 mag (AB) at 3.6 and 4.5$\mu$m. SEDS and S-CANDELS cover
smaller areas to even deeper limits.  We also make use of data from
SMUVS, which provides substantially deeper Spitzer/IRAC data over the deep 
UltraVISTA stripes.

Source catalogs were constructed using SExtractor v2.19.5 \citep{Bertin_1996}, run in dual image mode, with source detection performed
on the square root of a $\chi^2$ image \citep{Szalay_1999}
generated from the UltraVISTA YJHK$_{s}$ images.  In creating
our initial catalog of $z\sim7$ candidate galaxies, we started from
simply-constructed catalogs derived from the ground-based
observations.  Prior to our photometric measurements, images were first convolved to the
J-band point-spread function (PSF) and carefully registered against the
detection image (mean RMS$\sim$0.05\arcsec). Color measurements were made
in small \citet{Kron1980} like apertures (SExtractor AUTO and Kron factor
1.2) with typical radius $\sim$0.35-0.50\arcsec).

We also consider color measurements made in fixed 1.2$''$-diameter
apertures when refining our selection of $z\sim7$ candidate
galaxies. For the latter color measurements, flux from a source and its nearby neighbors (12\arcsec$\times$12\arcsec$\,$region) is carefully modeled; and then flux from the neighbors is subtracted before the aperture photometry is performed. Our careful modeling of the light from neighboring sources improves the overall robustness of our final
candidate list to source confusion. Total magnitudes are derived by correcting the fluxes measured in 1.2\arcsec-diameter apertures for the light lying outside a 1.2$"$-diameter aperture. The relevant correction factors are estimated on a source-by-source basis based on
the spatial profile of each source and the relevant PSF-correction kernel.  Photometry on the Spitzer/IRAC \citep{Fazio_2004} observations is more involved due to the much lower resolution FWHM = 1.7\arcsec$\,$compared to the ground-based data (FWHM = 0.7\arcsec). The lower resolution results in source blending where light from foreground sources contaminates measurements of the sources of interest. These issues can largely be overcome by modeling and subtracting the contaminating light using the higher spatial resolution near-IR images as a prior.  Measurements
of the flux in the {\it Spitzer}/IRAC observations were performed with the mophongo software \citep{Labbe_2006,Labbe_2010a,Labbe_2010b,Labbe_2013,Labbe2015}. The
positions and surface brightness distributions of the sources in the
coadded JHKs image are assumed to appropriate models and, after PSF
matching to the IRAC observations, these models are simultaneously fit
to the IRAC image leaving only their normalization as a free
parameter.  Subsequently, light from all neighbors is subtracted, and
flux densities were measured in 2\arcsec$\,$diameter circular apertures. The IRAC fluxes are corrected to total for missing light outside the
aperture using the model profile for the individual sources. The
procedure employed here is very similar to those of other studies
\citep[e.g.,][]{Skelton_2014,Galametz_2013,Guo_2013,Stefanon_2017,Weaver_2022}.

\begin{figure}[th!]
\epsscale{1.17}
\plotone{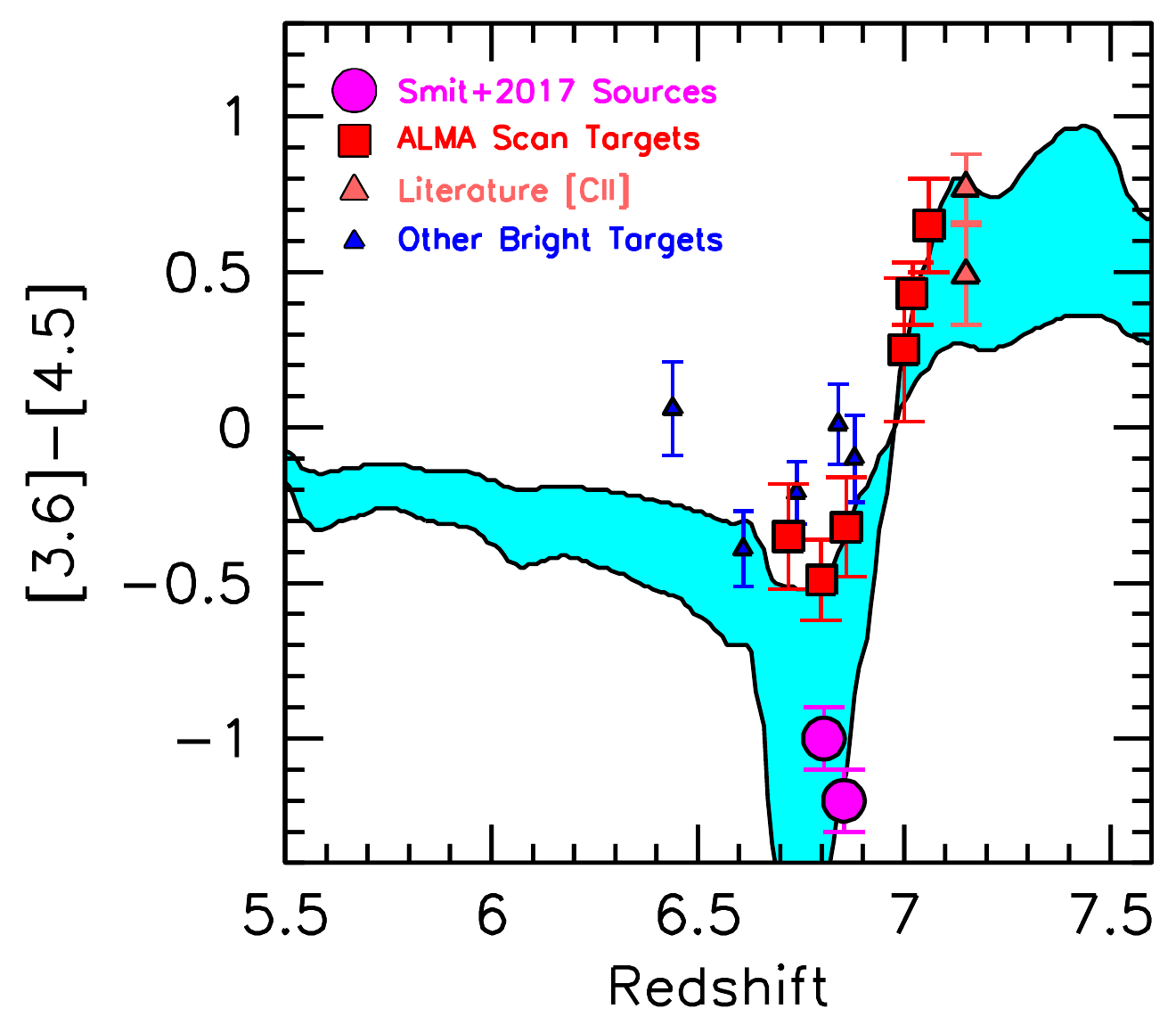}
\caption{Measured [3.6]$-$[4.5] colors for the bright $z\sim 7$ galaxy candidates we have identified within UltraVISTA.  The [3.6]$-$[4.5] color is largely driven by how high the EWs of the emission lines ([OIII]+H$\beta$, H$\alpha$) that fall in the [3.6] and [4.5] bands are, with the blue colors at $z\sim6.6$-6.9 driven by the [OIII]+H$\beta$ lines falling in the [3.6] band, and at $z>7$ the [OIII]+H$\beta$ lines falling in the [4.5] band.  The most promising $z\sim7$ follow-up targets are indicated by the red squares and correspond to the brightest $H\lesssim 25$ mag sources over COSMOS and show robustly red or blue [3.6]$-$[4.5] colors, significantly narrowing the width of the redshift likelihood distribution over which [CII] searches are required.  The filled magenta circles show the [3.6]$-$[4.5] colors measured for the two $z\sim 7$ galaxies with [CII] detections reported in \citet{Smit_2018Natur.553..178S}, while the solid orange triangles are for sources with [CII] detections in the literature \citep{Matthee_2017,Pentericci_2016,Hashimoto_2019_10.1093/pasj/psz049}. The blue triangles correspond to those bright ($H\leq 24.5$ mag) $z\sim7$ galaxies from Appendix A where the redshift is less well constrained based on the [3.6]$-$[4.5] colors (UVISTA-Z-002, 003, 004, 005, 008).  The cyan shaded region shows the expected [3.6]$-$[4.5] colors for star-forming galaxies vs. redshift assuming a rest-frame equivalent width for [OIII]+H$\beta$ in the range 400\AA to 2000\AA.}
\label{fig:z3645}
\end{figure}

\subsection{Bright $z\sim7$ Selection\label{sec:z7select}}

In effort to identify a robust set of $z\sim7$ galaxies from the wide-area UltraVISTA data set for follow-up with ALMA, we require sources to be detected at $>$6$\sigma$, combining the flux in J, H, Ks,
[3.6], and [4.5]-band images (coadding the S/N’s in quadrature).  The
combined UltraVISTA + IRAC detection and SNR requirements exclude
spurious sources due to noise, detector artifacts, and diffraction
features.  We construct a preliminary catalog of candidate z $\sim$ 7
galaxies using those sources that show an apparent Lyman break due to
absorption of UV photons by neutral hydrogen in the IGM blueward of
the redshifted Ly$\alpha$ line. This break is measured using simple
color criteria.  At $z > 6.2$, the $z$-band flux is significantly impacted by this absorption of rest-$UV$ photons, while at $z>7.1$, the $Y$-band flux is impacted.  The following criteria were applied:
\begin{displaymath}
(z-Y>0.5)\wedge(Y-K<0.7)
\end{displaymath}
where $\wedge$ is the logical AND operator.  In case of a
non-detection, the $z$ or $Y$-band flux in these relations is replaced
by the equivalent 1$\sigma$ upper limit.

It is worth emphasizing that our final sample of $z > 6$
bright galaxies shows little dependence on the specific limits chosen
here.  For each candidate source the redshift probability distribution
P(z) is then determined using the EAZY photometric redshift software \citep{Brammer_2008}, which fits a linear combination of galaxy
spectral templates to the observed spectral energy distribution
(SED).

The template set used here is the standard EAZY v1.0 template set,
augmented with templates from the Galaxy Evolutionary Synthesis Models \citep[GALEV:][]{Kotulla_2009} which include nebular continuum and
emission lines. The implementation of nebular lines follow the
prescription of \citet{Anders_2003}, assuming
0.2$\,$Z$_{\odot}$ metallicity and a large rest-frame EW of H$\alpha$ =
1300\AA.  These extreme EW reproduce the observed [3.6]$-$[4.5] colors
for many spectroscopically confirmed $z\sim7$-9 galaxies  \citep[e.g.,][]{Ono_2012,Finkelstein_2013,Oesch_2015,Zitrin_2015,Roberts_Borsani_2016,Stark_2017}.  A flat prior on the redshift is assumed.

To maximize the robustness of candidates selected for our $z\sim7$ samples, we require the integrated probability beyond $z = 6$ to be $>$70\%. The use of a redshift likelihood distribution $P(z)$ is very effective in rejecting faint low-redshift galaxies with a strong
Balmer/4000\AA$\,$break and fairly blue colors redward of the break.

The available optical observations are used to reject other low
redshift sources and Galactic stars by imposing $\chi^2 _{opt} < 4$.
$\chi^2 _{opt}$ is defined as $\chi_{opt} ^2 = \Sigma_{i}                                                
\textrm{SGN}(f_{i}) (f_{i}/\sigma_{i})^2$ where $f_{i}$ is the flux in
band $i$ in a consistent aperture, $\sigma_i$ is the uncertainty in
this flux, and SGN($f_{i}$) is equal to 1 if $f_{i}>0$ and $-1$ if
$f_{i}<0$.  $\chi^2 _{opt}$ is calculated in both 0.8$''$-diameter
apertures and in scaled elliptical apertures.  $\chi_{opt} ^2$ is
effective in excluding $z=1$-3 low-redshift star-forming galaxies
where the Lyman break color selection is satisfied by strong line
emission contributing to one of the broad bands \citep[e.g.][]{vanderWel_2011,Atek_2011}.  Sources which show a 2$\sigma$ detection in the
available ground-based imaging bands (weighting the flux in the
different bands according to the inverse square uncertainty in
$f_{\nu}$) are also excluded as potentially corresponding to
lower-redshift contaminants.  Finally, to minimize contamination by low-mass stars, we fit the observed SEDs of candidate $z\sim7$ galaxies both with EAZY and with stellar templates from the SpecX prism library \citep{Burgasser_2004}.  Sources which are significantly better
fit ($\Delta \chi^2$ $>$ 1) by stellar SED models are excluded. The SED templates
for lower mass stars are extended to 5$\mu$m using the known $J -[3.6]$ or
$J -[4.5]$ colors of these spectral types \citep{Patten_2006,Kirkpatrick_2011} and the nominal spectral types of stars from
the SpecX library. The approach we utilize is identical to the
SED-fitting approach employed by \citet{Bouwens_2015} for
excluding low-mass stars from the CANDELS fields.  Combined, these
selection requirements result in very low expected contamination
rates.

Using the above selection criteria, we select 32 $z\sim7$ potential candidates for spectral scans over a 1.2 square degrees area in the UltraVISTA field.  The sources have an H-band magnitude ranging from 23.8 mag to 25.7 mag and redshifts 6.6 to 7.1.  We include a list of the sources we select in Table~\ref{tab:candidates} from Appendix A.

\begin{deluxetable*}{lcccccc}[th!]
\tablecaption{Main parameters of the ALMA observations used for this study. \label{tab:tab1}}
\tablewidth{0pt}
\tablehead{\colhead{\specialcell[c]{Source \\ Name \\ }} & \colhead{\specialcell[c]{RA}} & \colhead{\specialcell[c]{DEC}}     & \colhead{\specialcell[c]{Beamwidth$^{a}$ \\ (arcsec)}} & \colhead{\specialcell[c]{Integration\\time$^{b}$\\(min.)} } & \colhead{\specialcell[c]{PWV \\ (mm)}} & \colhead{\specialcell[c]{Frequency/Redshift Range\\of the Spectral Scan\\Percentage of $p(z)^d$\\(GHz)}} }
\startdata
UVISTA-Z-001                                           & 10:00:43.36                                    & 02:37:51.3                                      & 1.47\arcsec$\times$1.21\arcsec                                   & 39.31                                                                  &3.3 &  228.62 - 239.37 ($z=6.94$-7.31) (71\%)                          \\
UVISTA-Z-007                                           & 09:58:46.21                                    & 02:28:45.8                                      & 1.40\arcsec$\times$1.19\arcsec                                   & 32.76                                                         &1.9& 240.28 - 251.02 ($z=6.57$-6.91) (82\%)           \\
UVISTA-Z-009                                           & 10:01:52.30                                    & 02:25:42.3                                      & 1.38\arcsec$\times$1.20\arcsec                                   & 32.76                                                      &1.9& 240.28 - 251.02 ($z=6.57$-6.91) (65\%)\\
UVISTA-Z-010                                           & 10:00:28.12                                    & 01:47:54.5                                      & 1.44\arcsec$\times$1.18\arcsec                                   & 39.31                                                      &3.3& 228.62 - 239.37 ($z=6.94$-7.31) (90\%)   \\
UVISTA-Z-013                                           & 09:59:19.35                                    & 02:46:41.3                                      & 1.45\arcsec$\times$1.18\arcsec                                   & 39.31                                                      &3.3& 228.62 - 239.37 ($z=6.94$-7.31) (99\%)    \\
UVISTA-Z-019                                           & 10:00:29.89                                    & 01:46:46.4                                      & 1.39\arcsec$\times$1.18\arcsec                                   & 32.76                                                      &1.9& 240.28 - 251.02 ($z=6.57$-6.91) (95\%)\\
\enddata
\tablecomments{
\textsuperscript{a} Beamsize for the naturally weighted moment-0 images.
\textsuperscript{b} Corresponds to the average on-source integration time for the two tunings. 
\textsuperscript{c} Average precipitable water vapour during the observations.
\textsuperscript{d} Percentage of the redshift probability distribution that is covered by the spectral scan.
}
\vspace{-0.5cm}
\end{deluxetable*}

\subsection{Target Selection for the ALMA Observations}

In an effort to further demonstrate the potential of spectral scans with ALMA to characterize massive star-forming galaxies at $z\sim7$, we elected to target six sources from the $z\sim7$ galaxy sample constructed in the previous Section (and which is presented in Appendix A).

We focused on those galaxies which are brightest in the rest-frame $UV$ and have the tightest constraints on their photometric redshifts.  $UV$-bright galaxies are particularly useful to target since those sources have the highest apparent SFRs and should contain particularly luminous [CII] lines, assuming the \citet{delooze2014} relation holds out to $z>6$ as \citet{Schaerer_2020} find.  If there is an additional contribution from obscured star formation, the luminosity of [CII] should be further enhanced.

Additionally, it is useful to target sources with tight constraints on their redshifts from photometry to minimize the number of spectral scan windows that need to be utilized.  For this purpose, a useful set of sources to target are those with particularly strong
constraints on their photometric redshifts from their \textit{Spitzer}/IRAC colors.  One particularly interesting opportunity exists for
sources where the broad-band Lyman-break places sources around a
redshift $z\sim7$, as \citet{Smit_2015} and \citet{Smit_2018Natur.553..178S} have already demonstrated.  This is due to the fact that at $z\sim7$, the
\textit{Spitzer}/IRAC color can reduce the width of the redshift likelihood
window for a source by as much as a factor of 2.  Due to the dramatic
impact the [OIII]+H$\beta$ lines have on the [3.6]$-$[4.5] colors for
star-forming galaxies at $z\sim7$, the \textit{Spitzer}/IRAC color places robust
constraints on the redshift of a source.  For sources with a robustly
red \textit{Spitzer}/IRAC [3.6]$-$[4.5] color, we can eliminate the $z<7$
solution, while for sources with a robustly blue \textit{Spitzer}/IRAC
[3.6]$-$[4.5] color, we can eliminate the $z>7$ solution.

Following from these arguments, the best targets for the detection of
luminous [CII] line emission at $z>6$ are those sources (1) which are
bright $(H<25)$, (2) have photometric Lyman breaks around $z\sim7$,
and (3) have robustly red or blue colors.  We highlight these sources
in a {\it Spitzer}/IRAC [3.6]$-$[4.5] color vs. redshift diagram in
Figure~\ref{fig:z3645} as the large red squares.

\begin{figure*}
\epsscale{1.18}
\plotone{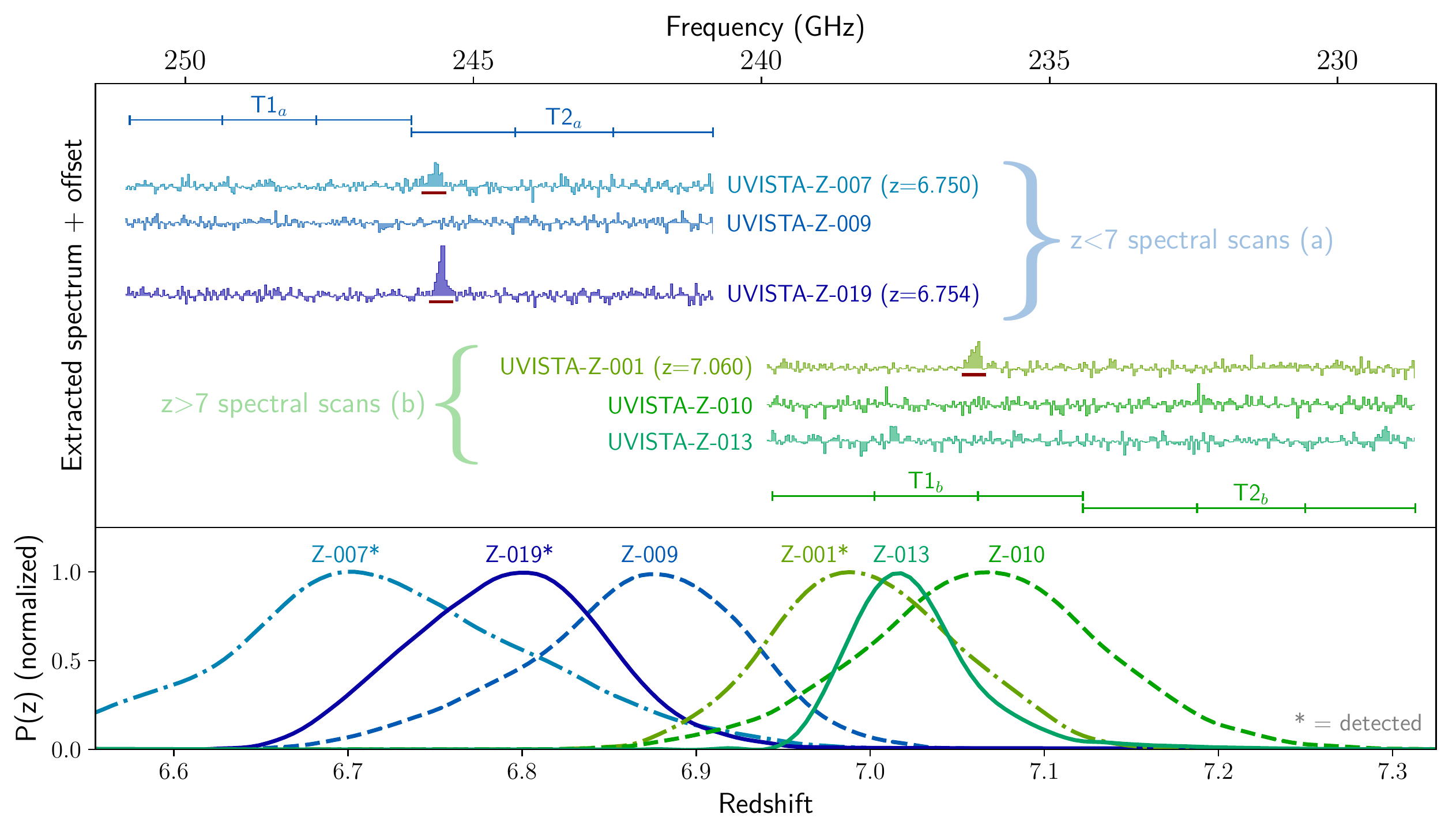}
\caption{\textit{(Top)} Full spectral scans as observed for the six galaxies targeted in this study. The targets are divided in two samples of three galaxies which were observed with different ALMA tunings. For both samples the ALMA tunings are shown above or below the spectral scans. We detect an emission line for 3 targets, indicated with a red line below the location of the emission line in the scans. For clarity, the spectra are binned combining 5 spectral channels. \textit{(Bottom)} Photometric redshift probability distributions for the galaxies targeted in this study (on the same scale as the spectral scans). Galaxies for which an emission line is detected are indicated with an asterisks. The two samples with different ALMA tunings are distinguished with blue and green colors.} \label{fig:scan}
\end{figure*}

\begin{deluxetable*}{lccccccccc}[]
\tablecaption{Results - UV and ALMA derived properties of the galaxies targeted in this study.\label{tab:results}}
\tablewidth{0pt}
\tablehead{\colhead{\specialcell[c]{Source\\Name}} & \colhead{\specialcell[c]{$z_{phot}$}} & \colhead{\specialcell[c]{$z_{spec}$}}     & \colhead{\specialcell[c]{$L_{UV}$$^\dagger$ \\ ($arcsec$)}} & \colhead{\specialcell[c]{$\log M_{*}$$^{\dagger}$\\$M_{\odot}$}} & \colhead{\specialcell[c]{EW OIII+H$\beta^{\ddagger}$\\($\AA$)}} & \colhead{\specialcell[c]{$L_{IR}$$^{a}$\\($10^{11}L_{\odot}$)}} & \colhead{\specialcell[c]{$S_{\nu}\Delta v$ \\ ($Jy$ $km/s$)}} & \colhead{\specialcell[c]{$L_{[CII]}$\\($10^8 L_{\odot})$}} & \colhead{\specialcell[c]{FWHM$^b$\\($km/s$)}}}
\startdata
UVISTA-Z-001 & 7.00$^{+0.05}_{-0.06}$ & 7.0599(3)& 2.9$^{+0.1}_{-0.1}$ & 9.58$^{+0.09}_{-0.35}$  & 1004$^{+442}_{-206}$   & 5.0$^{+2.1}_{-2.1}$ & 0.57$\pm$0.08   & 6.7 $\pm$ 1.2 & 256 $\pm$ 27 \\
UVISTA-Z-007 & 6.72$_{-0.09}^{+0.10}$ & 6.7496(5)& 1.5$^{+0.2}_{-0.2}$ & 9.57$^{+0.35}_{-0.44}$  & 761$^{+530}_{-168}$    & $<$ 2.2$^c$             & 0.51$\pm$0.09   & 5.6 $\pm$ 1.4 & 301 $\pm$ 42 \\
UVISTA-Z-009 & 6.86$_{-0.06}^{+0.07}$ & -        & 1.6$^{+0.2}_{-0.2}$ & 9.40$^{+0.32}_{-0.29}$  & 1012$^{+677}_{-257}$   & $<$ 2.4             & -             & -$^{d}$            & - \\
UVISTA-Z-010 & 7.06$_{-0.07}^{+0.07}$ & -        & 1.1$^{+0.2}_{-0.2}$ & 8.88$^{+0.28}_{-0.09}$  & 1706$^{+780}_{-807}$   & $<$ 2.1             & -             & -$^{d}$            & - \\
UVISTA-Z-013 & 7.02$_{-0.03}^{+0.03}$ & -        & 1.4$^{+0.4}_{-0.3}$ & 10.72$^{+0.03}_{-0.10}$ & 1821$^{+4364}_{-1142}$ & $<$ 2.2             & -             & -$^{d}$            & - \\
UVISTA-Z-019 & 6.80$_{-0.06}^{+0.05}$ & 6.7534(2)& 1.0$^{+0.1}_{-0.1}$ & 9.51$^{+0.19}_{-0.18}$  & 628$^{+226}_{-99}$     & 2.7$^{+0.9}_{-0.9}$ & 0.80$\pm$0.06 & 8.8 $\pm$ 0.9   & 184 $\pm$ 15 \\
\enddata
\tablecomments{
\textsuperscript{$\dagger$} $UV$-Luminosities and stellar masses are taken from \citet{Schouws_2021} and were derived using the methodology described in \citet{Stefanon2019}, assuming a metallicity of 0.2 $Z_{\odot}$, a constant star formation history and a \citet{Calzetti_2000} dust law.
\textsuperscript{$\ddagger$} [OIII]+H$\beta$ equivalent widths are taken from \citet{Bouwens_LP} and Stefanon et al. (2022, in prep).
\textsuperscript{a} Total infrared luminosity integrated from 8-1000$\mu$m assuming a modified black body SED with a dust temperature of 50$\,$K and a dust emissivity index $\beta_{dust}$=1.6 after correcting for CMB effects \citep[we refer to][for details]{Schouws_2021}.
\textsuperscript{b} Observed FWHM of the [CII] emission line as measured in the 1d spectrum.
\textsuperscript{c} UVISTA-Z-007 shows dust continuum emission at a level of $2.5\sigma$ (corresponding to $L_{IR}\sim1.8\times10^{11}L_{\odot}$), but we use the $3\sigma$ upper limit on the luminosity for the remainder of our analysis to be conservative.
\textsuperscript{d} These non-detections are either caused by a [CII] luminosity is lower than our detection limit ($\sim 2\times10^{8} L_{\odot}$, see Figure \ref{fig:efficiency}) or because their redshift falls outside of the range scanned in this study. Because of this degeneracy we cannot provide limits on their [CII] luminosity outside the scanned redshift ranges (see table \ref{tab:tab1}).
}
\end{deluxetable*}

\subsection{ALMA Observations and Data Reduction} \label{sec:datareduction}

A summary of the ALMA data collected for this second pilot program is presented in Table~\ref{tab:tab1}.  ALMA observations were obtained over a contiguous 10.75 GHz range (2 tunings) to scan for [CII] line.  For the three targets with photometric redshifts $z \lesssim 7$, the redshift range $z=6.57$ to 6.91 was scanned (240.28 to 251.02 GHz).  For the targets with photometric redshifts $z\gtrsim 7$, the redshift range $z=6.94$ to 7.31 was scanned (228.62 to 239.37 GHz).  The scan windows utilized are presented in Figure \ref{fig:scan}, along with the redshift likelihood distribution inferred from our photometry.  These scan windows cover between 71\% to 99\% of the estimated redshift likelihood distribution (Table~\ref{tab:tab1}).  The required integration times for the observations were set assuming a similar [CII] luminosity and line width for sources as in pilot program observations of \citet{Smit_2018Natur.553..178S}, i.e.,  $\sim$4$\times$10$^{8}$ $L_{\odot}$ and 200 km/s for the FWHM of [CII].  To detect this line at $>$5$\sigma$, we required achieving a sensitivity of 300 $\mu$Jy per 66 km/s channel.  To achieve this sensivitity, we require $\sim33$ to 39 minutes of integration time with ALMA.

The ALMA data were reduced and calibrated using \textsc{Casa} version 5.4.1 following the standard ALMA pipeline procedures. To reduce the data-size of the visibility measurement set, we averaged the data in bins of 30 seconds after carefully calculating that this bin-size does not impact our data through time-average smearing (e.g. Thompson, Moran \& Swenson 2017). We then performed initial imaging of the full data-cube using the \textsc{tclean} task with natural weighting. We clean to a depth of 3$\sigma$ per channel and use automasking\footnote{For the automasking we use the recommended settings for the 12-meter array with compact baselines from the \textsc{Casa} automasking guide: \url{https://casaguides.nrao.edu/index.php/Automasking\_Guide}. We verified that the automasking identifies the same emission regions that we would have selected when masking by hand.} to identify regions that contain emission. This initial data-cube was used to do an inital search for [CII] line candidates, details of our line-search procedure are described in the next Section.

If a significant emission line candidate is identified, we use the line properties to carefully mask the channels containing line emission to produce a continuum subtracted visibility data-set using the \textsc{uvcontsub} task.  This continuum subtracted measurement set is then used to re-image the full data cube, after which we repeat the line search and verify that the same line candidates are obtained.

\begin{figure*}[ht]
\epsscale{1.125}
\plotone{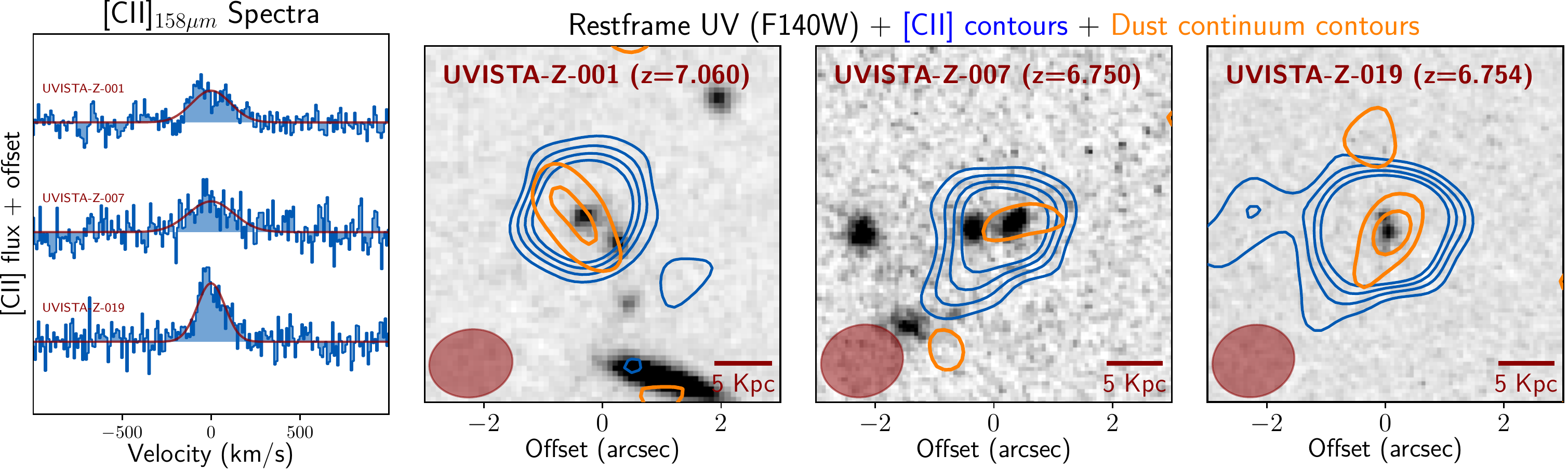}
\caption{Overview of the galaxies detected in [CII] in this study. (\textit{left panel}) 1d extracted spectra of the [CII] line (\textit{blue}) and a Gaussian fit (\textit{red}) for UVISTA-Z-001 (\textit{top}), UVISTA-Z-007 (\textit{middle}) and UVISTA-Z-019 (\textit{bottom}). (\textit{right panels}) The spatial distribution of the [CII] line emission (\textit{blue contours}) relative to rest-$UV$ images of the sources from HST (\textit{F140W}: \textit{background image}) and dust continuum emission (\textit{orange contours}).  Contours correspond to 2, 3, 4, and 5$\times$ the noise level. The dust continuum is significantly detected in UVISTA-Z-001 and UVISTA-Z-019 and marginally detected in UVISTA-Z-007 (at $2.5\sigma$) \citep[see also][for an extensive discussion of the continuum properties of these galaxies]{Schouws_2021}. 
\label{fig:morphology}} 
\end{figure*}

For each emission line candidate we produce an initial moment-0 image, including channels that fall within 2$\times$ the initial FWHM estimate of the line candidate\footnote{Collapsing over all channels that fall within 2$\times$ the FWHM captures $\sim$98\% of the flux for lines with a Gaussian line-profile.}. Using this moment-0 map, we produce a 1d spectrum where we include all pixel-spectra that correspond to $>3\sigma$ emission in the moment-0 map and we weigh the contribution of each pixel-spectrum by their signal-to-noise level. We then fit a Gaussian line model to this spectrum to extract the central frequency and the FWHM. Next, using this updated estimate for the FWHM, we update the moment-0 image and its associated signal-to-noise weighted 1d spectrum. We perform this 10 times and note that it converges to a stable solution within a few iterations. The line parameters that we derive with this method are also used to carefully exclude line emission from the continuum imaging used in \citet{Schouws_2021}.

\section{Results\label{sec:results}}

\subsection{[CII] Line Search} \label{sec:linesearch}
We performed a systematic search for emission line candidates using the MF3D line search algorithm \citep{Pavesi_2018}. MF3D finds line candidates using Gaussian template matching, which accounts for both spectrally and spatially extended emission lines. We used MF3D with 10 frequency templates with line-widths ranging from 50 to 600 km/s and 10 spatial templates ranging from 0 to 4.5 arcseconds. To be considered a reliable detection, we require line candidates to be within 1.5\arcsec$\,$of the rest-frame UV position of our sources and have SNR$>$5.2$\sigma$. This criterion was found to result in $>$95\% purity (Schouws et al. in prep.).

Based on this search, we find reliable emission lines for UVISTA-Z-001 at 12.8$\sigma$ at 235.80 GHz, for UVISTA-Z-007 at 9.4$\sigma$ at 245.24 GHz and for UVISTA-Z-019 at 18.3$\sigma$ at 245.12 GHz. The other datacubes did not contain any line candidates that meet the SNR requirements discussed above. For these non-detections, either their [CII] luminosity is lower than our detection limit ($\sim 2\times10^{8} L_{\odot}$, see Figure \ref{fig:efficiency}) or their redshift falls outside of the range scanned in this study. The results of the line-search are summarized in Figure \ref{fig:scan}, which shows the layout of the full spectral scan and corresponding P(z)'s for all six sources in this study.

For the detected sources we show the contours from the [CII] and dust continuum emission compared to the rest-frame $UV$ morphology and their 1d spectra in Figure \ref{fig:morphology}.  The rest-frame $UV$ images are in the F140W band at 1.39$\mu$m and are from GO-13793 \citep[UVISTA-Z-001, PI:Bowler:][]{bowler2017_10.1093/mnras/stw3296} and GO-16506 (UVISTA-Z-007, UVISTA-Z-019, PI: Witstok; Witstok et al. 2022, in prep).  For a detailed description of the procedure we used to produce the moment-0 images and 1d spectra we refer to Section \ref{sec:datareduction}.

We measure the integrated flux of the [CII] emission lines from the moment-0 images by fitting the data with a 1-component Gaussian model using the \textsc{imfit} task. The resulting flux measurements are shown in Table \ref{tab:results}. We double-check the measurements from \textsc{imfit} with \textsc{UVMULTIFIT} \citep{uvmultifit2014A&A...563A.136M}, which we use to fit a Gaussian model in the (u,v)-plane instead of the image-plane. We reassuringly find that both methods produce results consistent within their errorbars. Finally, we convert these [CII] fluxes to luminosities following \citet{Solomon_1992} and \citet{Carilli_2013}:
\begin{equation}
    L_{[CII]}/L_{\odot} = 1.04\times S_{\nu}\Delta\nu\times\nu_{obs}\times D_{L}^2
\end{equation}
with $S_{\nu}\Delta\nu$ the integrated flux density in mJy km/s, $\nu_{obs}$ the observing frequency and $D_{L}$ the luminosity distance.

\begin{figure}[t!]
\epsscale{1.25}
\plotone{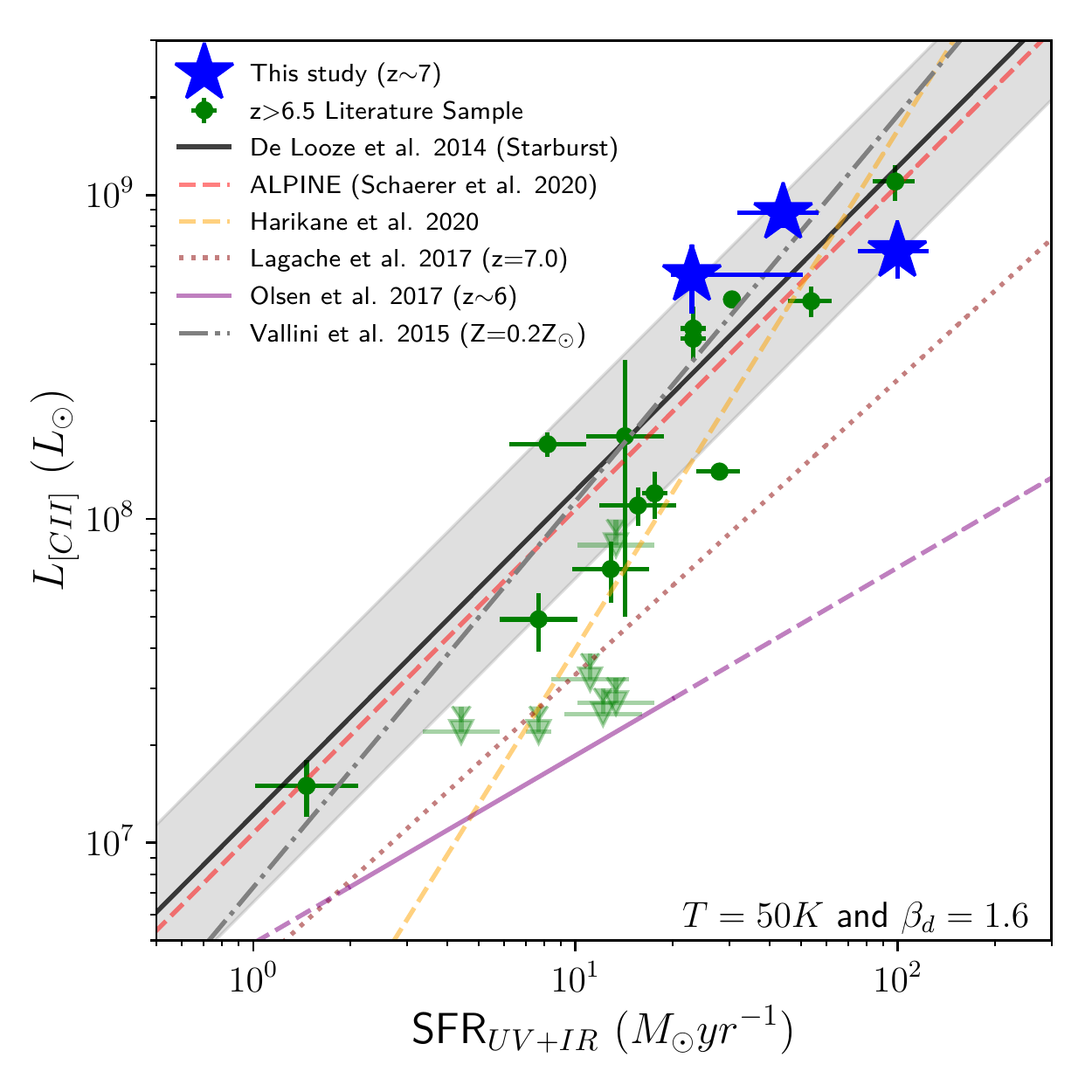}
\caption{[CII]-SFR relation for the galaxies in this study (\textit{solid blue stars}).  Our results are consistent with the results from \citet{delooze2014} for local HII/Starburst galaxies within the scatter (\textit{solid black line and grey shaded region}). For context we show results from previous z$>$6.5 detections and non-detections \citep[from the compilation by][]{Matthee_2019} (\textit{green data-points and upper-limits}). We also show some fits to the $L_{[CII]}$-SFR relation from the literature for observations \citep{Schaerer_2020,harikane_2020}, semi-analytic models \citep[][at $z=7$]{Lagache2018}, and zoom-in simulations by \citet{Olsen_2017} and \citet{Vallini_2015} (with $Z=0.2Z_{\odot}$). All SFRs have been scaled to a Chabrier IMF 
\citep{Chabrier_2003} and IR luminosities are calculated assuming a modified black body curve with $T=50K$ and $\beta_{d}=1.6$ \citep[as described in][]{Schouws_2021}.}\label{fig:SFR-CII} 
\end{figure}

\begin{figure}[t!]
\epsscale{1.25}
\plotone{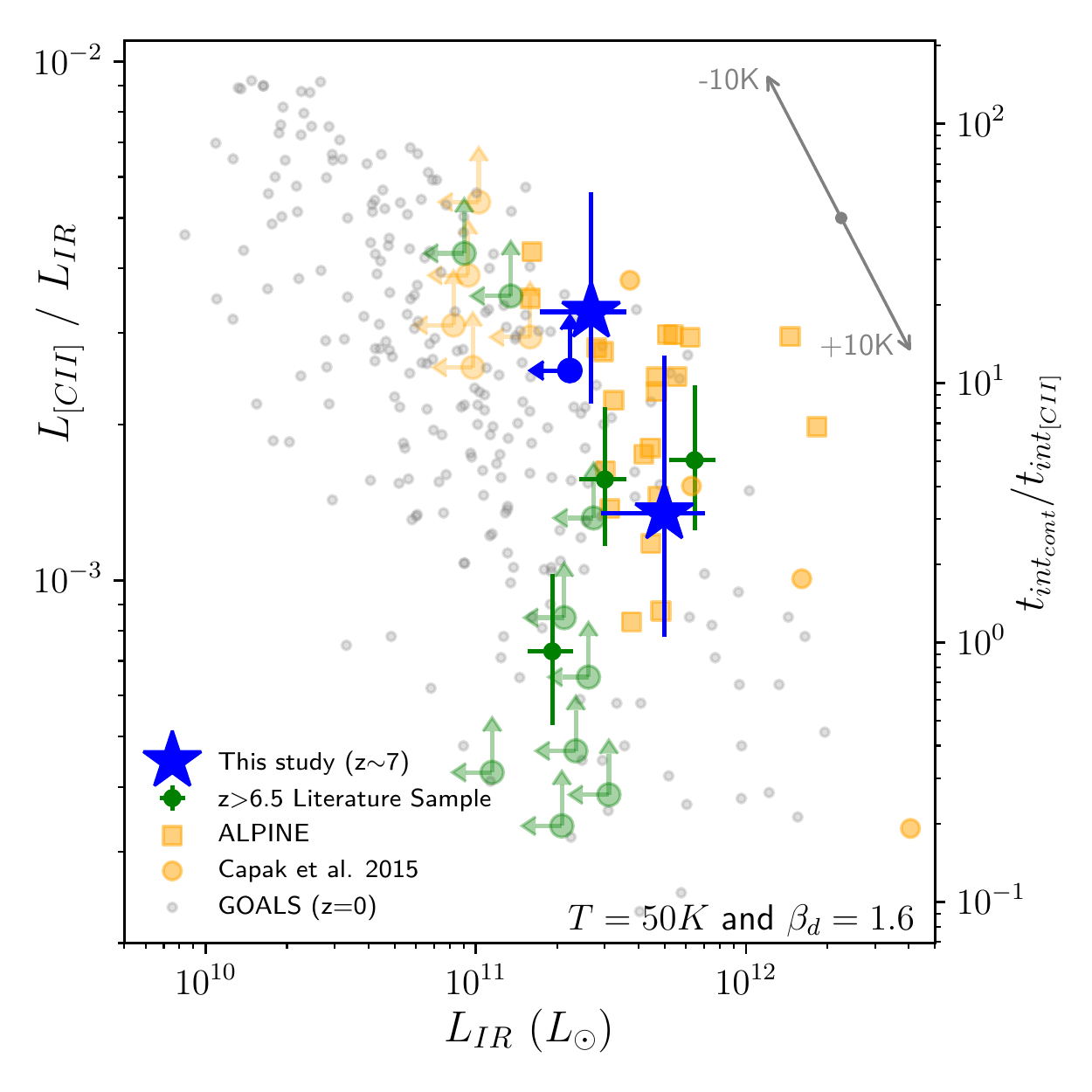} 
\caption{Ratio of the observed [CII] luminosity $L_{[CII]}$ to the IR luminosity $L_{IR}$ as a function of $L_{IR}$ for our small $z\sim7$ galaxy sample (\textit{solid blue stars}, $1\sigma$ uncertainties).  IR luminosities $L_{IR}$ are estimated assuming a modified blackbody SED with dust temperature 50K and an emissivity index $\beta_d$ of 1.6. With the grey arrows in the top-right corner we show the effect of changing the assumed dust temperature by $\pm$10K.  For context, we also show results from the $z=0$ GOALS sample \citep[][\textit{grey circles}]{Diazsantos2013,Diaz_Santos_2017}, the $z\sim4$-6 ALPINE sample \citep[][\textit{yellow squares}]{lefevre_2020,bethermin2020A&A...643A...2B,Faisst_2020_10.1093/mnras/staa2545}, and other results from the literature (\textit{green circles}).  Lower limits on $L_{[CII]}/L_{IR}$ and upper limits on $L_{IR}$ are $3\sigma$.  On the right vertical axis, the ratio of integration times required to detect [CII]$_{158}$ at $5\sigma$ to the time required to detect sources in the dust continuum at $3\sigma$ is indicated.  Our new measurements show slightly higher $L_{[CII]}/L_{IR}$ ratios than the $z=0$ GOALS results at a given $L_{IR}$ and appear to be qualitatively very similar to the $z\sim4$-6 results obtained by ALPINE.}\label{fig:CII-deficit}
\end{figure}

\begin{figure*}[th]
\epsscale{1.1}
\plotone{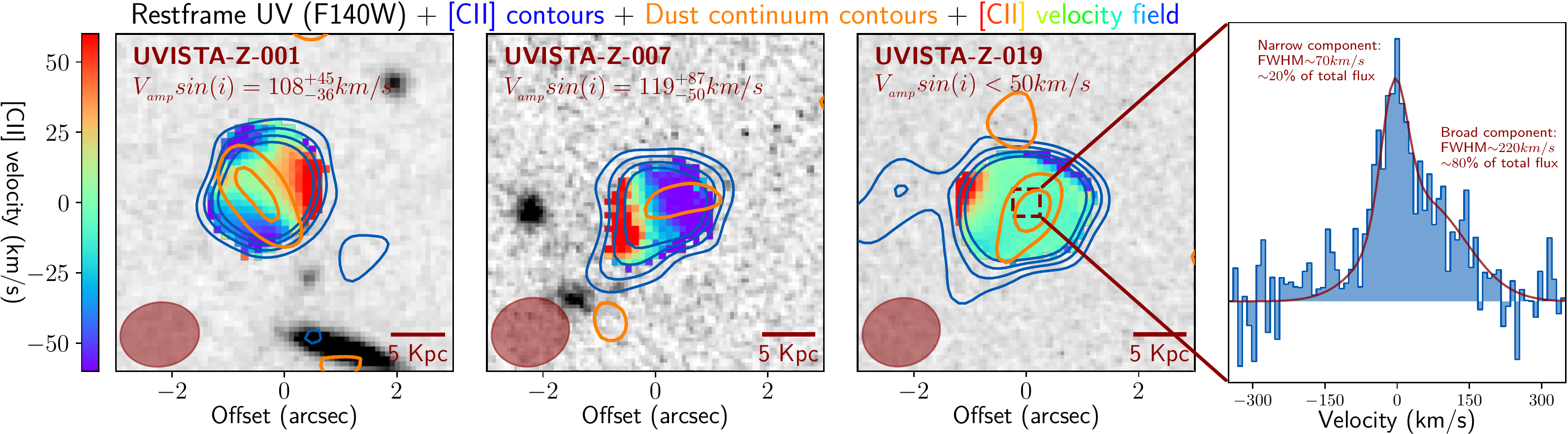}
\caption{(\textit{left three panels}) Low-resolution [CII] kinematics of our sources overlaid on the rest-UV imaging with the [CII] and dust continuum contours, as in Figure \ref{fig:morphology}.  The velocity gradients for all sources are shown on the same scale, starting at $-$60 km/s (\textit{blue}) to 60 km/s (\textit{red}). (\textit{rightmost panel}) Zoom-in on the central pixel-spectra of UVISTA-Z-019. The spectrum consists of a narrow component (FWHM $\sim70km/s$) responsible for $\sim20\%$ of the total flux and a broad component (FWHM $\sim220km/s$) that accounts for $\sim80\%$ of the total flux.} \label{fig:velocity}
\end{figure*}

\subsection{The [CII]-SFR Relation}
The large luminosity and favourable atmospheric opacity of [CII] enable detection up to very high redshifts.  Local galaxies studies have found a tight correlation between the [CII] luminosity and SFR \citep{deLooze_2011,delooze2014, Kapala_2014, cormier2015A&A...578A..53C, Herrera_Camus_2015, Diaz_Santos_2017}, [CII] has therefore been proposed as an efficient and unbiased probe of the SFR at high redshift. 

In past few years, this correlation between SFR and L$_{[CII]}$ has been observed out to $z\sim8$ with an increasing number of detections and upper-limits. Of particular note are the results from the ALPINE large program, which finds little to no evolution in the [CII]-SFR relation in a large sample of normal galaxies at $4.4<z<5.9$ \citep{Schaerer_2020}. At even higher redshifts the current samples are less uniform, but still seem to be consistent with the local relation, albeit with a larger scatter \citep[e.g.,][]{Carniani_2018,Matthee_2019}. 

Nevertheless, there has been an increasing number of observations of galaxies that fall well below the local relations
\citep[e.g.,][]{Ota_2014,Matthee_2019,Laporte201910.1093/mnrasl/slz094,Bakx2020,Binggeli2021A&A...646A..26B,Uzgil_2021,Rybak_2021,jolly2021}. In particular lensed galaxies probing lower M$_*$ and SFRs, and Lyman-$\alpha$ emitters (LAE) seem to be deficient in [CII] \citep{harikane_2020,jolly2021}.






We show the position of our sources on the [CII]-SFR relation in Figure \ref{fig:SFR-CII}, and find that the galaxies targeted in this study are also consistent with the local relation from \citet{delooze2014} within the expected scatter.  The three sources where [CII] remains undetected are not shown on this Figure, since it is unclear whether the [CII] line is below our detection threshold (i.e., $L_{[CII]} < 2 \times 10^{8} L_{\odot}$) or whether the true redshift is outside the range of our spectral scan. Because these sources have SFRs between 18 and 26 $M_{\odot}yr^{-1}$ our detection limit falls only slightly below and within the scatter from the local relation (\citet{delooze2014}). From this Figure, it is clear that most of the $z>6.5$ galaxies with SFR$\gtrsim 20 M_{\odot}yr^{-1}$ are consistent with the local relation, while at lower SFRs a significant fraction of currently observed galaxies seems to fall below the relation.

\subsection{[CII] vs. FIR}

It has been found that [CII] can account for up to $\sim1\%$ of the total infrared luminosity of galaxies, however it has also been found that this fraction decreases by $\sim2$ orders of magnitude with increasing $L_{IR}$, leading to a [CII]-deficit in luminous galaxies \citep[e.g.,][]{genzel2000,Malhotra_2001,Hodge2020arXiv200400934H}. Specifically, observed [CII]/FIR ratios range from $\sim$10$^{-2}$ for normal z$\sim$0 galaxies to $\sim$10$^{-4}$ for the most luminous objects, with a large scatter \citep[e.g.,][]{Diazsantos2013}.

The reason for this observed [CII]-deficit remains a topic of discussion in the literature, with a large range of possible explanations including optically thick [CII] emission, effects from AGN, changes in the IMF, thermal saturation of [CII], positive dust grain charging and dust-dominated HII region opacities \citep[e.g.,][]{Casey_2014b,Smith_2016,Ferrara2019,Rybak_2019,Hodge2020arXiv200400934H}.

The galaxies in this study have far infrared luminosities of $L_{IR} = 1-5\times 10^{11} L_{\odot}$\footnote{To calculate the far infrared luminosities we assume a modified black body dust-SED with $T=50$K and $\beta=1.6$, see \citet{Schouws_2021} for details.}, which puts them in the Luminous Infrared Galaxy (LIRG) classification. Comparing their infrared to their [CII] luminosity we find ratios of [CII]/$L_{IR}\sim1-3\times10^{-3}$ (see Figure \ref{fig:CII-deficit}). Compared to local (U)LIRGS from the GOALS survey \citep{Diazsantos2013,Diaz_Santos_2017}, we find that our galaxies are less deficient in [CII] by a factor $\sim$0.3 dex. This result has a minor dependence on the assumed dust temperature as shown with the grey arrows on Figure \ref{fig:CII-deficit}. When assuming higher or lower dust temperatures the data-points move mostly parallel to the trend. Only for substantially higher dust temperatures ($\gtrsim70$K) would our measurements be consistent with the local results.

Our measured [CII]/FIR ratios are consistent with other studies at high redshifts, which also find that high redshift galaxies tend to be less [CII] deficient \citep[e.g.,][]{Capak2015_Natur.522..455C,Schaerer_2020}. A possible explanation for this lack of [CII]-deficit in high redshift galaxies could be different dust conditions.   Specifically, a lower dust-to-gas ratio at a fixed far infrared luminosity could increase the [CII] luminosity with respect to the infrared \citep{Capak2015_Natur.522..455C}.


\subsection{[CII] Kinematics}
Due to its high intrinsic luminosity, [CII] is an efficient tracer of the kinematics of high redshift galaxies \citep[e.g.,][]{Neeleman_2019}. To investigate the kinematics of our sources, we derive the velocity maps of our galaxies by fitting Gaussians to the pixel-spectra in our cube, including all pixels for which the uncertainty on the velocity is less than 50 km/s. The resulting velocity fields are shown in Figure \ref{fig:velocity}. Despite the low ($\sim$1.3\arcsec, see Table \ref{tab:tab1}) resolution of our observations, some of our sources show significant velocity gradients. 

In particular, we find that both UVISTA-Z-001 and UVISTA-Z-007 display a significant velocity gradient, with velocity amplitudes of $v_{amp}\sin(i)=108^{+45}_{-36}$ km/s and $v_{amp}\sin(i)=119^{+87}_{-50}$ km/s respectively\footnote{We derive $v_{amp}\sin(i)$ by fitting a  rotating thin disk model to the 3D datacube using forward modeling with our kinematics fitting code \textsc{Skitter} (Schouws et al., in prep.). The maximum velocities on the kinematics maps shown in Figure \ref{fig:velocity} are lower than the actual $v_{amp}\sin(i)$ due to beam-smearing effects.}. If we compare the observed rotational velocities assuming no correction for inclination ($i=0$, hence $v_{obs}$ = $v_{amp}\sin(i)$) to the total line-width of the 1d spectrum ($\sigma_{obs}$) (see Table \ref{tab:results} and Figure \ref{fig:velocity}), we find that the $v_{obs}/\sigma_{obs}$ are $1.0_{-0.4}^{+0.6}$ and $0.9^{+0.9}_{-0.5}$ respectively. This would mean that both sources should most likely be classified as rotation dominated \citep[defined as $v_{obs}/\sigma_{obs}>0.8$ as utilized in e.g.][]{Forster_Schreiber_2009}. This calculation does not assume a correction for the inclination of the system, which would increase $v_{obs}/\sigma_{obs}$. The observed velocity gradient could however also be caused by close mergers. At the current resolution, rotating disks are indistinguishable from mergers \citep[e.g.][Schouws et al., in prep]{jones2020}.

In particular, we find indications that UVISTA-Z-007 could be a merger. The HST F140W imaging (Witstok et al. in prep.) shows clearly that this source consists of two distinct components (see Figure \ref{fig:morphology}). The observed velocity gradient is in the same direction as the two UV components (as shown in Figure \ref{fig:velocity}), making it likely that the observed velocity gradient in the [CII] is in fact due to the merger of these two components. 

For UVISTA-Z-019 we do not observe a significant velocity gradient and constrain the maximum rotation velocity to $v_{amp}sin(i)<50$ km/s, implying that this system is either dominated by dispersion or a face-on system ($i=0$). A more detailed look at the pixel-spectra within the cube indicates that in the central part of this source, the [CII] emission seems to break down into two distinct components (rightmost panel on Figure \ref{fig:velocity}), consisting of a narrow component (FWHM $\sim70$ km/s) responsible for $\sim20\%$ of the total flux and a broad component (FWHM $\sim220$ km/s) that accounts for $\sim80\%$ of the total flux. 

This interesting spectral feature could be caused by several processes, such as the effect of an outflow.  However, this would imply that the majority of the [CII] luminosity originates from the outflow and that the emission from the galaxy would have a very narrow FWHM, implying a low dynamical mass, lower than the stellar mass ($M_{dyn}\sim9\times10^{8}$ $M_{\odot}$\footnote{We derive dynamical masses following \citet{Wang_2013}: $\frac{M_{dyn}}{M_{\odot}sin(i)}=1.94\times10^{5}\cdot(\frac{FWHM}{km/s})^{2}\cdot \frac{r_{1/2}}{kpc}$ where $FWHM$ is the full width half maximum of the [CII] line in $km/s$ and $r_{1/2}$ the [CII] half light radius in kpc. Because the narrow [CII] component is unresolved, we assume a size of $r_{1/2}\sim$1 kpc \citep[consistent with][]{bowler2017_10.1093/mnras/stw3296}.} versus $M_{*}\sim3\times10^{9}$ $M_{\odot}$). An alternative explanation for the spectral feature could be a minor-merger, where the narrow component originates from an in-falling galaxy. However, this would mean that the LOS velocity is rather small at only $\Delta V \sim 50$ km/s, despite this likely being one of the final stages of the merger. Indeed, HST morphology tentatively shows two close components separated by $\sim$1.5 kpc (see also Figure \ref{fig:morphology}). Finally, the kinematics could also be evidence for a bright clump of intense star formation within a larger system, indicating the complex structure of high redshift sources \citep[e.g.,][]{kohandel2019,kohandel2020}. Hence an alternative interpretation of the clumps in HST imaging could be the presence of multiple star-forming regions in a larger system. Higher spatial resolution ALMA observations or deep rest-frame optical observations with JWST would be invaluable to definitively distinguish between these scenarios.

\begin{figure}[t!]
\epsscale{1.125}
\plotone{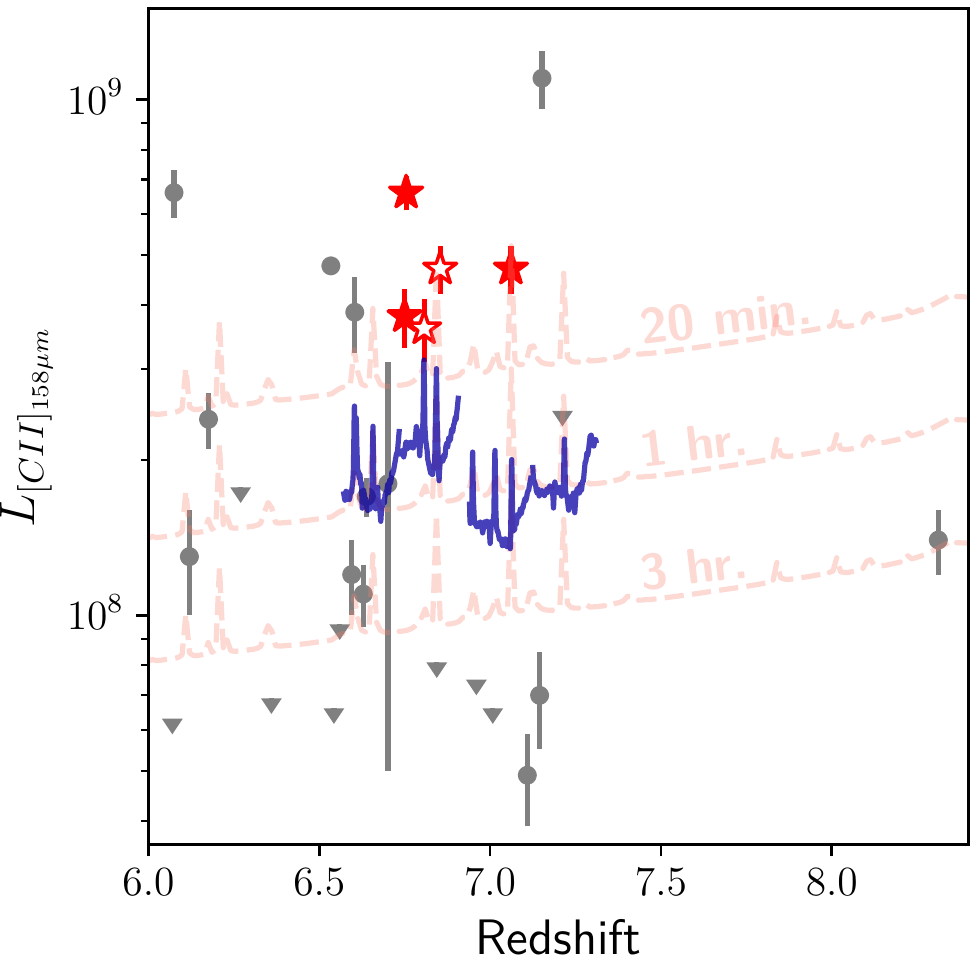} 
\caption{In this study we have presented a method to identify and efficiently spectroscopically confirm the redshifts of luminous [CII] emitting galaxies (\textit{solid (this study) and open \citep{Smit_2018Natur.553..178S} red stars}).  We compare the derived luminosities of our sources to the literature (\textit{grey datapoints and upper-limits}).  The present compilation is from \citet{Matthee_2019} and \citet{Bakx2020}.  The newly discovered lines are more luminous than most previous detections.  The luminosities of those sources without [CII] detections are less clear, but for those sources where our scans cover the full likelihood distribution, it is likely that the lines are fainter and more in line with the typical sources.  For context we show the expected limiting luminosities (for a 5$\sigma$ detection and 350 km/s FWHM) that can be achieved with ALMA with 20, 60 and 180 minute integrations (\textit{red dashed lines}) and find that our targets could have been detected in as little as 20 minutes per scan-window. We also indicate the achieved depth of our observations (30-40 minutes) with the \textit{solid blue lines}. This demonstrates the great potential to use ALMA for spectral scans to obtain spectroscopic redshifts of UV-luminous galaxies in the epoch of reionization.} \label{fig:efficiency}
\end{figure}

\section{The Efficiency of Spectral Scans and Future Prospects\label{sec:potential}}

In this study we have obtained redshifts for three galaxies without a prior spectrocopic redshift. In particular by targeting UV-luminous galaxies with high SFRs, the [CII] lines we detect are also luminous. In Figure \ref{fig:efficiency} we show that the [CII] emission of our sources could have been detected at $>$5$\sigma$ in only $\sim20$ minute integrations per tuning with ALMA.

Our sources benefit from very tight constraints on the photometric redshift due to the large break in the IRAC colors (see Figure \ref{fig:z3645}), which enables us to cover a significant fraction of the $P(z)$ with only 2 tunings. For sources that lack this additional constraint, 3 or 4 tunings would be necessary to cover the $P(z)$ appropriately. Nevertheless, this would still mean that galaxies like the ones targeted in this study could be spectroscopically confirmed in less than $<1.5$ hours per source.

It should be noted though that for sources with a lower SFR (and hence lower [CII] luminosities), spectral scanning becomes expensive quickly. A spectral scan targeting an $L_*$ galaxy ($SFR_{UV}\sim 8 M_{\odot}/yr$ at $z=7$) would cost $\sim12$ hours on source adopting the \citet{delooze2014} $L_{[CII]}$-SFR relation.  Therefore, to study the [CII] emission from $L\leq L_*$ galaxies, either targeting lensed galaxies for spectral scans or following-up galaxies with a prior spectroscopic redshift (e.g. from JWST or ground-based Lyman-$\alpha$) remain the most suitable options.

One significant advantage of using a spectral scan strategy is the time spent integrating in regions of the spectra not containing prominent ISM cooling lines.  These integrations allow us to probe continuum emission from our targets.  This is important -- since the continuum is much harder to detect than [CII], as one can see from the [CII]/$L_{IR}$ ratios of our sources.  This is illustrated in Figure \ref{fig:CII-deficit} (right vertical axis), on which we show the ratio of the integration time needed to detect the dust continuum versus the [CII] line. We find that it is necessary to integrate up to $\sim20\times$ longer to obtain a $3\sigma$ detection of the dust continuum. This means that the time spent observing tunings that do not contain a spectral line is not wasted, but contributes to the necessary sensitivity to detect the faint dust continuum.

Based on the results presented in this paper and \citet{Smit_2018Natur.553..178S}, we proposed and were awarded the time to apply the spectral scan method to a significantly larger sample of galaxies, covering a much larger range in galaxy properties and redshift. The result was the on-going Reionization Era Bright Emission Line Survey (REBELS) large program, in which we pursue spectral scans for [CII] or [OIII] in a sample of 40 $z>6.5$ galaxies \citep{Bouwens_LP}.

\section{Summary}

In this paper we present the results of new ALMA spectral scan observations targeting [CII] in a small sample of six luminous Lyman Break Galaxies at 
$z\sim 7$.  The targeted sources were identified from deep, wide-area near-IR, optical, and \textit{Spitzer}/IRAC observations and are particularly luminous.  The targeted sources also feature tight constraints on their redshifts leveraging the abrupt changes that occur in the IRAC colour around $z\sim7$ (where strong line emission from [OIII]+H$\beta$ shifts from the [3.6] to [4.5] band).  This improves the efficiency of the spectral scans by $\sim$2$\times$ based on a small number of tunings required to cover the inferred $P(z)$. The present results build on the exciting results from \citet{Smit_2018Natur.553..178S}, who previously demonstrated the potential of spectral scans for [CII] with just two sources.

Our main results are summarized below:

\begin{itemize}
\item We detect ($>9\sigma$) [CII] lines for three of the six galaxies we target with our spectral scans (shown in Figure \ref{fig:scan}). The [CII] lines are strong with luminosities between $5.6 \times 10^{8}L_{\odot}$ and $8.8\times10^{8}L_{\odot}$. We also observe that the [CII], dust and rest-frame UV emission are well aligned within the resolution of our observation (see Figure \ref{fig:morphology}).

\item Placing our new detections on the [CII]-SFR relation shows that our sources are consistent with the local relation from \citet{delooze2014} (for HII/starburst galaxies) (shown in Figure \ref{fig:SFR-CII}), and we find slightly higher [CII]/$L_{IR}\sim1-3\times10^{-3}$ compared to local (U)LIRGS (see Figure \ref{fig:CII-deficit}), which is consistent with previous studies of high redshift galaxies.

\item Although our observations are taken at a relatively low resolution ($\sim$1.3\arcsec), we find that our sources display a broad spectrum of kinematic diversity. One of our sources seems to be rotation dominated, one source is most likely a major merger and one source is dominated by dispersion. We also find possible kinematic evidence for a bright star forming clump within the dispersion dominated source (see Figure \ref{fig:velocity}). However, higher resolution observations are necessary to confirm our interpretation of the kinematics of our sources.

\item We discuss the lack of evolution of the [CII]-SFR relation found for luminous high redshift galaxies by reviewing the literature on the physical effects that drive the [CII] emission in high redshift galaxies. While one would naively expect a trend towards lower [CII]/SFR values with redshift based on the higher ionization parameter, lower metallicities and higher densities of high redshift galaxies, this is not observed. We speculate that a lower dust-to-gas or dust-to-metal ratio, which increases the [CII] emission, could compensate for those effects.
\end{itemize}

These new results illustrate the tremendous potential spectral scans with ALMA have for characterizing luminous galaxies in the epoch of reionization (see Figure \ref{fig:efficiency}), including deriving spectroscopic redshifts for sources, and probing the kinematics and dynamical masses of sources, as well as the dust continuum \citep{Schouws_2021}.  Results from this data set showed the potential (\S\ref{sec:potential}) and were important in successfully proposing for the REBELS large program in cycle 7 \citep{Bouwens_LP}.  Future studies (Schouws et al. 2022, in prep) will significantly add to the current science using that considerable data set.\\ \\

\acknowledgements
\subsection*{Acknowledgements}
We are greatly appreciative to our ALMA program coordinator Daniel Harsono for support with our ALMA program.  This paper makes use of the following ALMA data: ADS/JAO.ALMA 2018.1.00085.S.  ALMA is a partnership of ESO (representing its member states), NSF (USA) and NINS (Japan), together with NRC (Canada), MOST and ASIAA (Taiwan), and KASI (Republic of Korea), in cooperation with the Republic of Chile. The Joint ALMA Observatory is operated by ESO, AUI/NRAO and NAOJ. Sander Schouws and Rychard Bouwens acknowledge support from TOP grant TOP1.16.057 and a NOVA (Nederlandse Onderzoekschool Voor Astronomie) 5 grant. JH acknowledges support of the VIDI research programme with project number 639.042.611, which is (partly) financed by the Netherlands Organisation for Scientific Research (NWO). HSBA acknowledges support from the NAOJ ALMA Scientific Research Grant Code 2021-19A. JW and RS acknowledge support from the ERC Advanced Grant 695671, "QUENCH", and the Fondation MERAC. RS acknowledges support from an STFC Ernest Rutherford
Fellowship (ST/S004831/1). The PI acknowledges assistance from Allegro, the European ALMA Regional Center node in the Netherlands.

\newpage
\appendix

\section{Bright Sample of $z\sim7$ Galaxies Used for Selecting Our Six Follow-up Targets}

For completeness, we provide the coordinates, magnitudes, [3.6]$-$[4.5] colors, and photometric redshifts for the full selection of bright $z\sim7$ candidate galaxies we identified (\S\ref{sec:z7select}) within the UltraVISTA field in Table~\ref{tab:candidates}.  \\ \\ 

\begin{deluxetable*}{cccccccc}[hb]
\tablewidth{0pt}
\tablecolumns{8}
\tabletypesize{\footnotesize}
\tablecaption{Candidate $z\sim7$ galaxies in the UltraVISTA DR3 observations\label{tab:candidates}}
\tablehead{
\colhead{ID} &
\colhead{R.A.} &
\colhead{Dec} &
\colhead{$m_{AB}$\tablenotemark{a}} &
\colhead{[3.6]-[4.5]} &
\colhead{$z_{phot}$\tablenotemark{b}} &
\colhead{$z_{spec}$} &
\colhead{References\tablenotemark{c}}}
\startdata
\multicolumn{8}{c}{Current Selection of Bright Candidate $z\sim7$ Galaxies}\\
UVISTA-Z-001  &  10:00:43.361  &  2:37:51.33 & 23.8$\pm$0.1 & 0.3$\pm$0.2 &7.00$_{-0.06}^{+0.05}$  & & [2]\\ 
UVISTA-Z-002  &  10:02:06.469  &  2:13:24.18 & 24.1$\pm$0.1 & $-$0.2$\pm$0.1 &6.74$_{-0.06}^{+0.06}$ & & [1,10] \\ 
UVISTA-Z-003  &  10:02:06.701  &  2:34:21.42 & 24.2$\pm$0.1 & $-$0.1$\pm$0.1 &6.88$_{-0.08}^{+0.08}$ \\ 
UVISTA-Z-004  &  10:01:36.850  &  2:37:49.10 & 24.3$\pm$0.1 & 0.0$\pm$0.1 &6.84$_{-0.09}^{+0.09}$ & & [2,10]\\ 
UVISTA-Z-005  &  10:01:58.501  &  2:33:08.22 & 24.3$\pm$0.1 & $-$0.4$\pm$0.1 &6.61$_{-0.10}^{+0.10}$ & & [1,10]\\ 
UVISTA-Z-006  &  10:01:40.688  &  1:54:52.37 & 24.4$\pm$0.1 & 0.5$\pm$0.2 &7.09$_{-0.05}^{+0.06}$ & 7.152$^{*}$ & [2,9]\\ 
UVISTA-Z-007  &  09:58:46.214  &  2:28:45.75 & 24.4$\pm$0.1 & $-$0.4$\pm$0.2 &6.72$_{-0.09}^{+0.10}$ \\ 
UVISTA-Z-008  &  09:58:39.762  &  2:15:03.27 & 24.4$\pm$0.1 & 0.1$\pm$0.2 &6.44$_{-0.05}^{+0.05}$ \\ 
UVISTA-Z-009  &  10:01:52.304  &  2:25:42.27 & 24.5$\pm$0.1 & $-$0.3$\pm$0.2 &6.86$_{-0.06}^{+0.07}$ & & [2]\\ 
UVISTA-Z-010  &  10:00:28.121  &  1:47:54.47 & 24.5$\pm$0.1 & 0.6$\pm$0.2 &7.06$_{-0.07}^{+0.07}$ & & [2]\\ 
UVISTA-Z-011  &  10:00:42.125  &  2:01:57.10 & 24.6$\pm$0.1 & $-$0.1$\pm$0.1 &6.55$_{-0.06}^{+0.05}$ & & [2,10]\\ 
UVISTA-Z-013  &  09:59:19.353  &  2:46:41.31 & 24.8$\pm$0.2 & 0.4$\pm$0.1 &7.02$_{-0.03}^{+0.03}$ & & [10]\\ 
UVISTA-Z-014  &  09:57:35.723  &  1:44:56.40 & 24.8$\pm$0.1 & --$^{d}$ &7.13$_{-0.16}^{+0.15}$ \\ 
UVISTA-Z-015  &  10:00:23.772  &  2:20:36.98 & 24.8$\pm$0.1 & 0.8$\pm$0.1 &7.07$_{-0.03}^{+0.03}$ & 7.154$^{**}$ & [2,3,4,5,6,7] \\ 
UVISTA-Z-016  &  10:01:54.562  &  2:47:35.79 & 24.9$\pm$0.2 & $-$1.1$\pm$0.9 &6.77$_{-0.26}^{+0.24}$ \\ 
UVISTA-Z-017  &  10:00:30.188  &  2:15:59.71 & 24.9$\pm$0.1 & $-$1.3$\pm$0.2 &6.78$_{-0.04}^{+0.04}$ & 6.854$^{***}$ & [1,2,3,4,8,10]\\ 
UVISTA-Z-018  &  10:02:03.811  &  2:13:25.06 & 25.0$\pm$0.2 & $-$0.6$\pm$0.3 &6.81$_{-0.10}^{+0.12}$ & & [2]\\ 
UVISTA-Z-019  &  10:00:29.892  &  1:46:46.37 & 25.0$\pm$0.2 & $-$0.5$\pm$0.1 &6.80$_{-0.06}^{+0.05}$ \\ 
UVISTA-Z-020  &  10:01:57.140  &  2:33:48.76 & 25.0$\pm$0.2 & $-$0.6$\pm$0.3 &6.73$_{-0.11}^{+0.12}$ \\ 
UVISTA-Z-021  &  09:57:36.994  &  2:05:11.28 & 25.1$\pm$0.2 & $-$0.3$\pm$0.3 &6.73$_{-0.16}^{+0.16}$ & & [10]\\
UVISTA-Z-022  &  09:59:13.206  &  2:21:52.51 & 25.1$\pm$0.2 & $-$0.1$\pm$0.6 &6.52$_{-0.16}^{+0.14}$ \\ 
UVISTA-Z-023  &  09:58:45.961  &  2:39:05.94 & 25.1$\pm$0.2 & $-$3.4$\pm$7.3 &6.76$_{-0.13}^{+0.13}$ \\ 
UVISTA-Z-024  &  09:59:04.558  &  2:11:38.10 & 25.2$\pm$0.2 & $-$0.8$\pm$0.4 &6.89$_{-0.14}^{+0.15}$ \\ 
UVISTA-Z-025  &  09:58:49.216  &  1:39:09.70 & 25.3$\pm$0.3 & $-$1.6$\pm$0.6 &6.70$_{-0.12}^{+0.12}$ & & [10]\\ 
UVISTA-Z-026  &  10:02:05.967  &  2:06:46.13 & 25.3$\pm$0.2 & $-$1.4$\pm$0.4 &6.74$_{-0.06}^{+0.06}$ & & [10]\\ 
UVISTA-Z-027  &  09:59:22.426  &  2:31:19.45 & 25.3$\pm$0.2 & 0.1$\pm$0.7 &6.69$_{-0.11}^{+0.11}$ & & [10]\\ 
UVISTA-Z-028  &  10:00:54.819  &  1:50:05.28 & 25.3$\pm$0.2 & $-$0.1$\pm$0.2 &6.67$_{-0.16}^{+0.18}$ & & [10]\\ 
UVISTA-Z-029  &  10:00:41.097  &  2:29:31.13 & 25.3$\pm$0.2 & $-$0.1$\pm$0.1 &6.72$_{-0.17}^{+0.16}$ \\ 
UVISTA-Z-030  &  10:02:22.458  &  2:04:45.72 & 25.3$\pm$0.3 & $-$0.1$\pm$0.2 &6.61$_{-0.10}^{+0.11}$ \\ 
UVISTA-Z-031  &  09:59:01.407  &  2:28:02.13 & 25.4$\pm$0.2 & $-$0.9$\pm$0.4 &6.61$_{-0.23}^{+0.19}$ \\ 
UVISTA-Z-032  &  10:00:22.482  &  1:45:32.62 & 25.4$\pm$0.2 & $-$1.0$\pm$0.4 &6.71$_{-0.17}^{+0.11}$ 
\enddata
\tablecomments{
\textsuperscript{a} UVISTA $H$-band magnitude.
\textsuperscript{b} 68\% confidence intervals derived by EAzY using the SED template set presented in \S\ref{sec:z7select}.
\textsuperscript{c} References: [1] \citet{Tilvi_2013}, [2] \citet{Bowler_2014,bowler2017_10.1093/mnras/stw3296}, [3] \citet{Bouwens_2015}, [4] \citet{Smit_2015}, [5] \citet{Roberts-Borsani_2016}, [6] \citet{Pentericci_2016}, [7] \cite{Stark_2017}, [8] \citet{Smit_2018Natur.553..178S}, [9] \citet{Hashimoto_2019_10.1093/pasj/psz049}, [10] \citet{Endsley_2021}.
\textsuperscript{d} This source fell outside the footprint of the IRAC mosaics used for the source selection in this manuscript.
\textsuperscript{*} Detected in dust continuum, [CII] and [OIII] by \citet{Hashimoto_2019_10.1093/pasj/psz049} and dust continuum by \citet{Bowler_2018_10.1093/mnras/sty2368} (Big Three Dragons/ B14-65666).
\textsuperscript{**} Detected in [CII] by \citet{Pentericci_2016} and Lyman-$\alpha$ by \citet{Stark_2017}.
\textsuperscript{***} Detected in [CII] by \citet{Smit_2018Natur.553..178S}.
}
\end{deluxetable*}

\newpage
\bibliography{references}

\end{document}